\documentclass[reprint,aps,prl,superscriptaddress,nobalancelastpage,nofootinbib]{revtex4-2}

\usepackage{graphicx}
\usepackage{svg}
\usepackage{float}
\usepackage[dvipsnames]{xcolor}
\usepackage{amsmath,amssymb}

\definecolor{darkblue}{rgb}{0,0,0.6}
\definecolor{darkred}{rgb}{0.6,0,0}
\usepackage[colorlinks=true,urlcolor=darkblue, citecolor=darkblue, linkcolor=darkred, hyperfootnotes=false]{hyperref}

\begin{document}

\title{Enhanced viscous adhesion using deformable structure}

\author{Mary Williams}
\affiliation{Laboratoire Interfaces $\&$ Fluides Complexes, Universit\'e de Mons (UMons), 20 Place du Parc, B-7000 Mons, Belgium.}
\author{Thomas Desmedt}
\affiliation{Universit\'e libre de Bruxelles (ULB), Nonlinear Physical Chemistry Unit, CP231, 1050 Bruxelles, Belgium.}
\author{Fabian Brau}
\email[]{fabian.brau@ulb.be}
\affiliation{Universit\'e libre de Bruxelles (ULB), Nonlinear Physical Chemistry Unit, CP231, 1050 Bruxelles, Belgium.}
\author{Pascal Damman}
\email[]{pascal.damman@umons.ac.be}
\affiliation{Laboratoire Interfaces $\&$ Fluides Complexes, Universit\'e de Mons (UMons), 20 Place du Parc, B-7000 Mons, Belgium.}

\date{\today}

\begin{abstract}
We investigate the adhesion dynamics of a thin elastic structure in contact with a viscous fluid and retracted at a controlled speed, mimicking natural adhesion mechanisms. During detachment, the viscous fluid confined between the deformable structure and a rigid substrate generates an adhesive force due to a pressure drop within the thin film. We show from dedicated experiments that the structural flexibility introduces a strongly nonlinear mechanical response, which significantly alters both the magnitude and the evolution of the adhesion force with retraction velocity. In contrast to rigid systems, the deformability of the structure enables enhanced and tunable adhesion. To capture this interplay, we develop a theoretical framework that couples elasticity and viscosity, providing new insights into how flexible structures enable adhesion control.
\end{abstract}

\maketitle

Nature offers powerful paradigms for functional adhesion, where animals combine high adhesion for contact and low adhesion for detachment to achieve effective locomotion. Evolution has produced amazing adaptations, such as the gecko pads coated with thousands of micrometric setae branched into nanometric spatulas, allowing for unprecedented dry adhesion through van der Waals forces~\cite{Autumn2002}. In contrast, tree frogs use smooth pads coated with mucus, which allow attachment to smooth, rough, and dry substrates~\cite{Federle2006,langowski2018tree}. Some insects, such as flies and beetles, have also developed complex pads decorated with hundreds of hairs secreting tiny droplets of viscous liquid~\cite{walker1985adhesive,gorb1998design,eisner2000defense,dirks2011fluid}. 

In these biological systems, adhesion involves complex interactions between soft or flexible structures and a rigid substrate through a fluid layer. Fluid-mediated adhesion is driven by a depression generated by capillary forces~\cite{PGG2004} and/or viscous forces~\cite{bikerman1947fundamentals}. The impact of the flexibility of the structure on adhesion has essentially been studied in the quasistatic regime, where capillary forces dominate~\cite{Gernay2016,Butler2019,gilet2019adhesive,Duprat2020}. In this case, the equilibrium adhesion force can be significantly enhanced due to the deformability of the structure~\cite{Butler2019,Duprat2020}. When the structure is removed at a finite speed from the substrate, viscous stresses may overcome capillary forces since their strength is proportional to the withdrawal speed and the fluid viscosity~\cite{bikerman1947fundamentals}. Interestingly, these two quantities are not small in the living realm. Indeed, the viscosity of pad secretion in beetles and ants is about 100 times that of water~\cite{Federle2002,Abou2010} and the retraction speed of the pad in beetles is about 2 mm/s~\cite{Iazzolino2020}. Despite its potential interest in the locomotion of some animals, the impact of the flexibility of the contacting microstructures on viscous adhesion is still poorly understood, even though it is known that it affects the adhesive strength~\cite{Poivet2004,Brau2016,Houze2017}.

Here we demonstrate, through combined experimental and theoretical approaches, the major influence of the microstructures morphology on adhesion strength by studying a model system based on the Elastica. Specifically, we measure the force allowing the retraction at constant speed of a flexible ring separated from a rigid substrate by a thin layer of viscous fluid. We show, in contrast to previous studies focusing on the rheology of the adhesive fluid~\cite{Linghu2025,Jensen2026}, that the adhesion properties can be controlled by the nonlinear elastic response of the contactor itself. The high deformability of the rings produces a gain in the adhesion strength that can be as large as two orders of magnitude. This elasto-viscous adhesion results from the interplay between the elastic force, which governs the deformation of the ring, and the viscous force, which fixes the imposed displacement at which the ring detaches from the substrate. We develop a theoretical model that couples these two effects to shed light on the resulting complex mechanical behavior.

\begin{figure*}
\centering
\includegraphics[width=\textwidth]{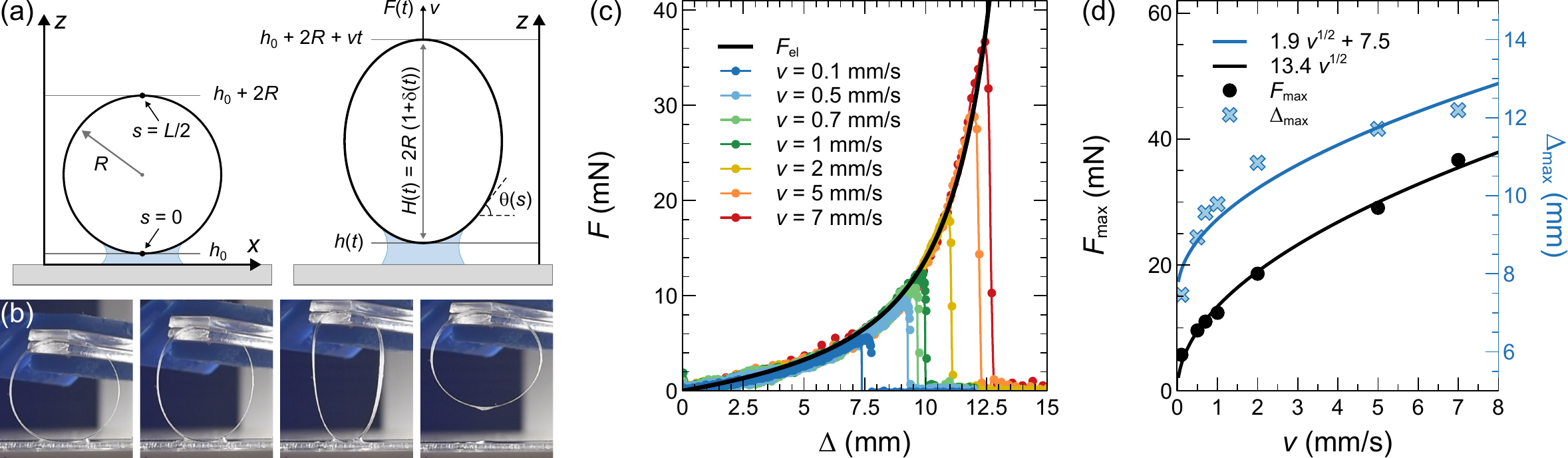}
\caption{(a) Schematic of the system showing the initial state of a PET ring of thickness $t$ and radius $R$ attached to a rigid substrate with a layer of fluid of thickness $h_0$ and viscosity $\mu$. The upper part of the ring is attached to a traction device and is moved upward with a constant speed $v$ (and thus a varying force $F$). This induces a relative elongation $\delta(t) = (H(t) - 2R)/2R$ of the ring and a variation $h(t)-h_0$ of the fluid thickness along the $z$-axis. (b) Snapshots of an experiment with a PET ring ($t=23$ $\mu$m, $R=14$ mm) in contact with silicon oil ($\mu = 1$ Pa s) and retracted at $v = 1$ mm/s. (c) Adhesion force as a function of the imposed displacement $\Delta$ for a PET ring ($t = 23$ $\mu$m and $R = 14$ mm) retracted at various speeds ($v = 0.1-7$ mm/s) together with the nonlinear elastic force $F_{\text{el}}$ (\ref{eq:fin}). The sudden drop of the adhesion force corresponds to the breaking of the viscous layer between the ring and the substrate. The point at which this drop occurs defines $F_{\text{max}}$ and $\Delta_{\text{max}}$. (d) Evolution of $F_{\text{max}}$ and $\Delta_{\text{max}}$ as a function of the retraction speed for the same flexible ring together with power law fits.}
\label{fig:manip}
\end{figure*}

\textit{Methods---}The experiments are performed with Polyethylene terephthalate (PET) rings of width $W = 15$~mm, thickness $23 \le t \le 350$ $\mu$m, and radius $8 \le R \le 22$ mm that have been prepared with a conventional LASER cutter. This range of thickness leads to a variation of the bending modulus $B= Et^3/12(1-\nu^2)$ of about three orders of magnitude ($10^{-5} \lesssim B \lesssim 10^{-2}$ Nm) where $E$ is the Young modulus and $\nu$ is the Poisson ratio. The rings are attached to a traction device (Instron) that measures the force $F$ required to vertically displace their upper parts at a constant retraction speed of $0.1 \le v \le 7$ mm/s [Fig.~\ref{fig:manip}(a),(b)]. To measure viscous adhesion, a droplet of silicon oil with a volume of $\Omega \simeq 100$~$\mu$L, viscosity $\mu \simeq 1$ Pa s and density $\rho \simeq 970$ kg/m$^{3}$, is placed between the bottom part of the rings and a rigid plate. The initial thickness of the fluid $h_0$ ranges between $400-800\mu$m.

\textit{Results---}Figure~\ref{fig:manip}(c) shows that the measured adhesive force increases with the imposed displacement $\Delta = v t$ until it abruptly drops to zero at detachment. The increasing part of the force–displacement curves obtained at various retraction speeds collapses onto a well-defined envelope, characterized by a linear regime at small displacements, followed by a pronounced supra-linear response at larger displacements. The maximum adhesive force, $F_{\text{max}}$, and the corresponding displacement, $\Delta_{\text{max}}$, at which detachment occurs, increase with the retraction speed as $\sqrt{v}$ [Fig.~\ref{fig:manip}(d)]. This nonlinear behavior indicates that a simple force balance between elastic, $F_{\text{el}} \sim \Delta$, and viscous contributions, $F_{\mu} \sim v$, cannot describe the observed mechanical response even in the small force regime. This discrepancy highlights the intrinsically coupled and rate-dependent nature of the response, which requires a deeper analysis. To address this problem, we first decouple it by considering two quasi-independent processes: (i) a purely elastic mechanical response of the rings subjected to a retraction force, which should describe the envelope of the adhesive force observed in Fig.~\ref{fig:manip}(c); (ii) the viscous flow between two plates (curved or not) that ultimately leads to detachment. 
By coupling these elastic and viscous components, we derive a theoretical model that describes the complete nonlinear mechanical response of these systems.

\textit{Mechanical response of a flexible ring---}Figure~\ref{fig:manip}(c) shows that the adhesion force increases with the retraction speed and that the system behaves as if the ring is effectively glued to the substrate before the sudden detachment. The elastic response of the ring is described through the Elastica equilibrium equation for half of the system due to the symmetry
\begin{equation}
   BW \frac{d^2\theta}{ds^2} + \frac{F_{\text{el}}}{2} \cos\theta = 0, 
\label{eq:origine}
\end{equation}
where $0\le s \le L/2$ is the arc length, $\theta(s)$ the angle between the tangent to the curve and the $x$-axis, $W$ the width of the ring, and $B$ its bending modulus~\cite{Landau1986}. The boundary conditions (BCs) are $\theta(0)=0$ and $\theta(L/2)=\pi$, where the length $L$ of the ring determines the initial undeformed radius $R=L/2\pi$ [Fig.~\ref{fig:manip}(a)]. With the dimensionless variable $s = \bar{s}\, L/2$, Eq.~(\ref{eq:origine}) becomes
\begin{equation}
    \frac{d^2\theta}{d\bar{s}^2} + q^2 \cos\theta=0, \ \text{with } q^2 = \frac{F_{\text{el}} L^2}{8BW} \equiv \frac{\pi^3}{4} \bar{F}_{\text{el}},
\label{eq:depart}
\end{equation}
with the BCs, $\theta(0)=0$ and $\theta(1)=\pi$. For each value of $\bar{F}_{\text{el}}>0$, the solution of Eq.~(\ref{eq:depart}) is a function $\theta(\bar s)$ from which the relative elongation along the $z$-axis is computed: $\delta=(H-2R)/2R$ with $H=\int_0^{L/2} \sin \theta(s) ds$ (see Supplemental Material~\cite{sup}). Figure~\ref{fig:glue} shows the numerical evolution of $\bar{F}_{\text{el}}$ as a function of $\delta$ obtained with Eq.~(\ref{eq:depart}) together with the shape of the ring at various $\delta$. The force grows linearly for small deformations before increasing strongly when $\delta$ approaches the full extension limiting value $\delta_{\text{max}}=\pi/2-1$. Because this elastic force will be used to model the fluid-structure interaction reported in Fig.~\ref{fig:manip}, it is useful to obtain an analytical expression of $\bar{F}_{\text{el}}$.
We use an asymptotic method that addresses both small (linear) and large (nonlinear) deformations. Equation~(\ref{eq:depart}) can be linearized around the circular solution to obtain~\cite{sup}
\begin{equation}
    \bar{F}_{\text{el}} \equiv \frac{2 R^2 F_{\text{el}}}{\pi B W} \simeq 8.56\, \delta, \quad \delta \ll 1.
\label{eq:small}
\end{equation}
Figure~\ref{fig:glue} shows that this asymptotic expression agrees well with the numerical data. 

The elastic force in the large deformation limit can be obtained by considering that the ring adopts a strong anisotropic shape composed of two semi-circles of radius $r$ and two straight segments of length $\mathcal{H}$ when $\delta \lesssim \delta_{\text{max}}$ (see inset Fig.~\ref{fig:glue}). These two lengths can be related to the radius of the ring $R$ and the relative elongation $\delta$. For this purpose, we note that inextensibility imposes that $\mathcal{H}=\pi (R-r)$. Next, the vertical elongation of this shape is $\mathcal{H}+2r$, so that the relative elongation is given by $\delta = (\mathcal{H}+2r - 2R)/2R$. Substituting the expression of $\mathcal{H}$ and solving for $r$, we find 
\begin{equation}
\label{r}
r(\delta) = R \, [\pi - 2(1+\delta)]/(\pi-2).
\end{equation}
All the bending energy $E_B$ of this idealized shape is stored in the two semi-circles of curvature $\kappa = 1/r$ and is therefore given by
\begin{equation}
\label{EB}
    E_B(\delta) = (B/2) \int \kappa^2 ds =\pi BW/r(\delta).
\end{equation}
The elastic force is obtained from the variation of the energy when the displacement is varied $F_{\text{el}} = {dE_B}/{d\Delta} = 1/(2R) {dE_B}/{d\delta}$:
\begin{equation}
    \bar{F}_{\text{el}} = \frac{2\alpha(\pi-2)}{[\pi-2(1+\delta)]^2},
    \label{eq:div}
\end{equation}
where a numerical factor $\alpha = 0.765$ is added to accurately fit the numerical solution of $\bar{F}_{\text{el}}$. We note that Eq.~(\ref{eq:div}) diverges when $\delta = \delta_{\text{max}} = \pi/2-1$ as it should. Finally, by adding the two asymptotic expressions (\ref{eq:small}) and (\ref{eq:div}) and removing their common part, i.e. the expansion of Eq.~(\ref{eq:div}) to the first order in $\delta$, we construct the global elastic force: 
\begin{equation}
    \bar{F}_{\text{el}}  \simeq 3.86\, \delta + \frac{1.747}{[\pi-2(1+\delta)]^2} -1.34.
\label{eq:fin}
\end{equation}
where the numerical constants have been obtained by fitting. Figure~\ref{fig:glue} shows that the experimental data overlap when the measured force is rescaled according to Eq.~(\ref{eq:depart}) and $\Delta$ by $2R$. The data are in excellent agreement with the theoretical elastic force given by Eq.~(\ref{eq:fin}) that also describes very well the envelope of the experimental force-displacement curves shown in Fig.~\ref{fig:manip}(c).

\begin{figure}[!t]
\centering
\includegraphics[width=\columnwidth]{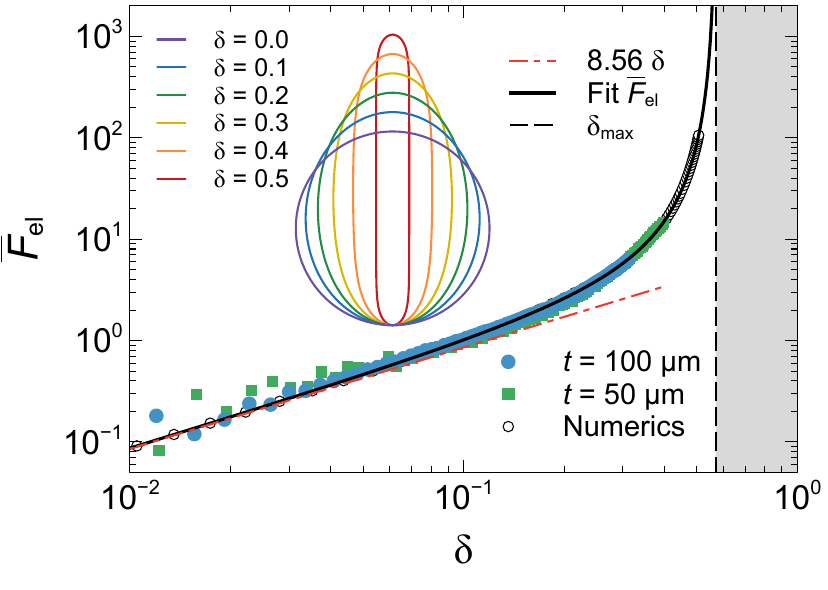}
\caption{Comparison between the theoretical normalized force-displacement curve for a ring obtained from the solution of Eq.~(\ref{eq:depart}) and experimental data for two flexible rings glued to a rigid substrate and retracted at $v = 1$ mm/s. The black solid curve shows the nonlinear relationship (\ref{eq:fin}) and the red dashed-dotted curve shows the linear regime (\ref{eq:small}) valid at small deformations. The shaded area shows the region $\delta > \delta_{\text{max}}=\pi/2-1$ that cannot be reached with an inextensible ring. The inset shows the shapes of the ring for different values of $\delta$.}
\label{fig:glue}
\end{figure}

\textit{Viscous force between two rigid plates in relative motion---}The viscous force required to separate two rigid plates of width $W$ at a speed $v$, between which lies a volume $\Omega$ of viscous fluid of thickness $h(x)$, can be calculated using the lubrication approximation~\cite{sup}. In 2D, the relationship between $v$, $h(x)$ and the pressure $p$ is given by the Reynolds equation,
\begin{equation}
   12 \mu v = \frac d{dx} \left[ h^3(x)\frac{dp}{dx}\right],
   \label{eq_Reynolds}
\end{equation}
with $dp/dx|_{x=0}=0$ and $p(\ell/2)=0$, where $-\ell/2 \le x \le \ell/2$ is the region where the fluid is located. This equation can be integrated twice to obtain
\begin{equation}
  p(x) = 12 \mu v \int_{\ell/2}^x \frac{x'}{h^3(x')} dx'.
\end{equation}
For two rigid flat plates of width $W$, we have $h(x)=h_0$ which yields the force of adhesion,
\begin{equation}
\label{eq_FAflat}
    F_\mu^\text{flat}(h_0) = -2W \int_{0}^{\ell/2} p(x) dx = \frac{\mu v\, \Omega^3}{W^2 h_0^6},
\end{equation}
where $\Omega = W \ell h_0$ is the volume of fluid. 

For a curved moving plate approximated by a circular arc with a radius $R$, the adhesion force and fluid volume can be obtained by the integrals~\cite{sup}
\begin{subequations}
    \label{eq_FAOmega2}
\begin{align}
    F_\mu&= 24\mu v W \int_0^{\ell/2} \frac{x^2}{h^3(x)}\,  dx,\\
    \Omega&= 2W \int_0^{\ell/2} h(x)\,  dx,
\end{align}
\end{subequations}
with $h(x) = R+ h_0 - \sqrt{R^2-x^2}$. Equations (\ref{eq_FAOmega2}) are the parametric equations for the adhesion force as a function of the volume of liquid, with $\ell$ as the parameter. This parametric curve can be well approximated by the following expression~\cite{sup}
\begin{equation}
    F_\mu(h_0,R)= \frac {F_\mu^\text{flat}(h_0)}{\left\{1+\left[F_\mu^\text{flat}(h_0)/F_{\mu}^{\text{R}}(R/h_0)\right]^{1/2}\right\}^2}.
    \label{eq_FACurved}
\end{equation}
The numerator is the force for a flat moving plate, and the denominator is the correction due to the curvature of the moving plate with $F_{\mu}^{\text{R}}=(6\pi/\sqrt 2) \mu v W (R/h_0)^{3/2}$. With this relation, $F_\mu = F_\mu^\text{flat}$ when $R/h_0 \to \infty$ and $F_\mu = F_{\mu}^{\text{R}}$ for $R/h_0 \to 0$. 

\begin{figure}[t!]
\centering
\includegraphics[width=\columnwidth]{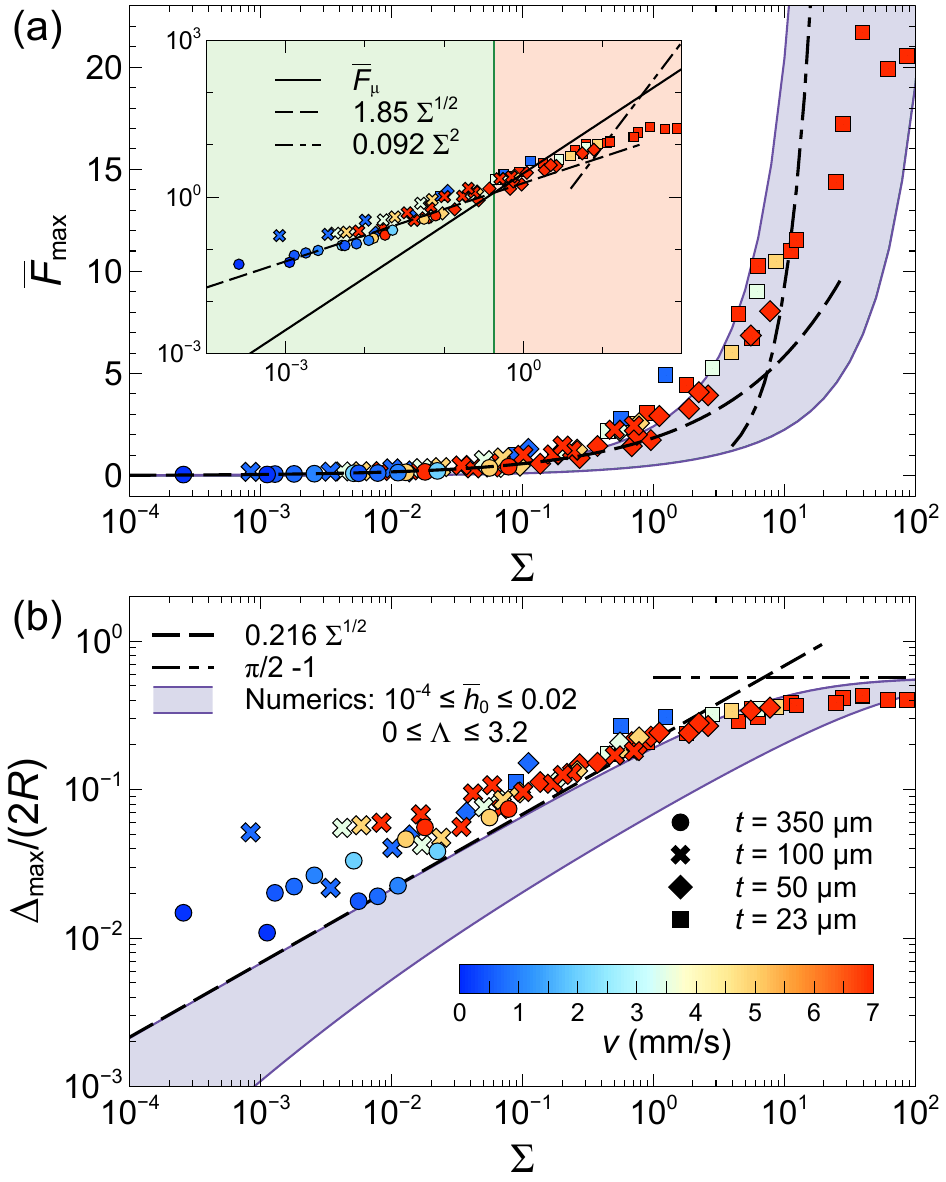}
\caption{(a) Normalized maximum force recorded when the rings detach from the substrate as a function of $\Sigma$ for various ring thicknesses and retraction speeds. The dashed and dashed-dotted curves represent, respectively, the asymptotic scalings (\ref{asymp-scaling-Sigma-small}) and (\ref{eq:grandsigma}) valid for $\Lambda=0$ and $\bar{h}_0 \ll 1$. The purple area shows the region spanned by $\bar{F}_{\text{max}}$ obtained from Eq.~(\ref{equ:nonlinear_ODE}) when $\Lambda$ and $\bar{h}_0$ vary respectively between $0$ and $3.2$ and between $10^{-4}$ and $0.02$. The upper limit of these interval correspond to the average experimental value of these parameters. Inset: same graph in log-log scale where the adhesive force between rigid structures, $\bar{F}_{\mu} = \Sigma/[\bar{h}_0 (1+\Lambda)^2]$, has been added for $\bar{h}_0 = 0.02$ and $\Lambda= 3.2$. (b) Same as panel (a) for the maximum normalized imposed displacement recorded when the rings detach from the substrate.}
\label{fig:fluide}
\end{figure}

\textit{Elasto-viscous adhesion model---}To model the global adhesion process, the viscous force should be combined with the mechanical response of the flexible ring. Neglecting inertia, the dynamics is given by $F_{\mu}(h,R) = F_{\text{el}}(\delta)$ where $F_{\text{el}}$ and $F_{\mu}$ are given, respectively, by Eqs.~(\ref{eq:fin}) and (\ref{eq_FACurved}) with $v=dh/dt$. To use this equation, we must relate the elastic deformation of the ring, $\delta(t)$ to the fluid thickness, $h(t)$. This is achieved by noting that the upper part of the ring is at a vertical position $z=h_0+2R+vt$, see Fig.~\ref{fig:manip}(a). This position results from the sum of the thickness of the fluid, $h(t)$, and the length of the elongated axes of the deformed ring, $H(t)$. From the definition of $\delta$, we have $H(t)= 2R (1+\delta(t))$, which yields $2R\delta(t) = h_0+vt - h(t)$. With the dimensionless variables, $\bar h=h/h_0$, $\bar h_0=h_0/2R$, $\bar t= vt/2R$, the dynamics of adhesion are governed by the nonlinear ODE,
\begin{subequations}
\label{equ:nonlinear_ODE}
\begin{align}
\label{equ:nonlinear_ODE-1}
&\frac{\Sigma}{\bar h^6 \left[1 + \Lambda \,\bar{h}^{-9/4}\right]^2}
\frac{d\bar{h}}{d\bar{t}}
=\bar{F}_{\text{el}}\left[\bar{h}_{0}(1-\bar{h}) + \bar{t}\right], \\
&\Sigma = \frac{\mu v R \Omega^3}{\pi BW^3 h_0^5}, \quad \Lambda = \frac{2^{1/4}}{\sqrt{6\pi}}\, \left[\frac{\Omega}{W h_0^{3/2} R^{1/2}}\right]^{3/2}.
\end{align}
\end{subequations}
The model involves three parameters: the normalized initial fluid thickness $\bar h_0$ together with $\Sigma$ and $\Lambda$ combining all the relevant physical parameters. $\Sigma$ essentially compares the characteristic viscous (\ref{eq_FAflat}) and elastic (\ref{eq:small}) forces. It can be viewed as a normalized retraction velocity that takes into account the ability of the flexible ring to store the imposed deformation. $\Lambda$ is a geometric parameter encoding the impact of the ring curvature on adhesion. In the experiments, $\bar h_0$ and $\Lambda$ explore a rather limited range, while $\Sigma$ varies over 5 orders of magnitude. With the proposed dimensionless variables, all the experimental data thus collapse onto a single master curve, $\bar F_\text{max}(\Sigma)$ (Fig.~\ref{fig:fluide}).

Asymptotic analytical expressions can be obtained in small $\Sigma$ and large $\Sigma$ regimes. From the temporal evolution of the fluid thickness $\bar{h}(\bar{t})$ as well as the adhesion force $\bar{F}_{\text{el}}$~\cite{sup}, it appears that i) the fluid thickness remains almost constant and suddenly increases at detachment, ii) the adhesion force follows the elastic response of the glued rings up to the sudden drop in force at detachment. Within the assumption of negligible curvature and considering that $h \simeq h_0$ before detachment, we have $\delta=\bar h_0 (1-\bar h) + \bar t \simeq \bar t$ and Eq.~(\ref{equ:nonlinear_ODE-1}) simplifies to $\bar{h}^{-6} (d\bar{h}/d\bar{t})\ \Sigma \simeq \bar{F}_{\text{el}} (\bar{t})$ which can be integrated with the initial condition $\bar{h}(0)=1$ to give the approximate solution,
\begin{equation}
\label{approx-solution-h}
    \bar{h}(\bar{t}) = {\Sigma}^{1/5} \left[\Sigma - 5\, G_{\text{el}}(\bar{t}) \right]^{-1/5},
\end{equation}
where $G_{\text{el}} = \int_0^{\bar{t}} \bar{F}_{\text{el}}(t') dt'$, with $\bar{F}_{\text{el}}$ given by Eq.~(\ref{eq:fin}). The approximate solution Eq.~(\ref{approx-solution-h}) describes the numerical solutions of Eq.~(\ref{equ:nonlinear_ODE}) well~\cite{sup}. The detachment occurs when $\bar h(\bar t)$ diverges, i.e. when  $G_{\text{el}}(\bar{t}_{\text{max}}) = \Sigma/5$. When $\Sigma$ is small, the detachment occurs very rapidly in the linear regime of the elastic force $\bar{F}_{\text{el}} \simeq 8.56\, \bar{t}$. After integration, we obtain $G_{\text{el}}\simeq 4.28\, \bar{t}^{\,2}$ which finally yields the maximum displacement and force
\begin{equation}
\label{asymp-scaling-Sigma-small}
    \bar{F}_{\text{max}} \simeq 1.85\, \Sigma^{1/2}, \quad \frac{\Delta_{\text{max}}}{2R} \simeq 0.216\, \Sigma^{1/2}, \quad \Sigma \ll 1.
\end{equation}
These relations predict that both $F_{\text{max}}$ and $\Delta_{\text{max}}$ vary as $\sqrt{v}$, in agreement with the results presented in Fig.~\ref{fig:manip}(d) and \ref{fig:fluide}. When $\Sigma$ becomes large, detachment occurs for an imposed displacement close to the divergence of the elastic force. Expanding $G_{\text{el}}$ near $\bar{t}=\pi/2-1$ and solving the detachment condition in the limit of large $\Sigma$ leads to the relations,
\begin{equation}
\bar{F}_{\text{max}} \simeq 0.092\, \Sigma^{2}, \quad
\frac{\Delta_{\text{max}}}{2R} \simeq \frac{\pi}{2}-1, \quad \Sigma \gg 1.
\label{eq:grandsigma}
\end{equation}
The theoretical variations of $\bar{F}_{\text{max}}$ and $\Delta_{\text{max}}$ with the rescaled retraction speed, ${\Sigma}$, are compared with a large set of experimental data in Fig.~\ref{fig:fluide}. The purple areas show the region spanned by the numerical values of the maximum displacement and force, obtained by solving the nonlinear ODE (\ref{equ:nonlinear_ODE}), when the fluid thickness $h_0$ and the curvature parameter $\Lambda$ vary. The asymptotic expressions (\ref{asymp-scaling-Sigma-small}) and (\ref{eq:grandsigma}) are also shown (valid when $\bar{h}_{0} \ll 1$ and $\Lambda=0$). The agreement between theory and experiments is generally good. 

The experimental data for the maximum force shown in Fig.~\ref{fig:fluide}(a) at large $\Sigma$ are significantly smaller than the asymptotic expression (\ref{eq:grandsigma}) obtained at $\Lambda=0$. This shows that the curvature of the deformable structure reduces the adhesion strength at large $\Sigma$. However, the inset of Fig.~\ref{fig:fluide}(a) shows that the deformability of the structure strongly increases adhesion at low $\Sigma$ compared to the rigid case (up to two orders of magnitude). Indeed, the maximum force for a rigid structure is given by Eq.~(\ref{eq_FACurved}) which reads $\bar{F}_{\mu} = \Sigma/[\bar{h}_0 (1+\Lambda)^2]\sim \Sigma$ and is smaller than $\bar{F}_{\text{max}}\sim \Sigma^{1/2}$ when $\Sigma$ is small enough. Therefore, the enhancement of adhesion force due to flexibility occurs for retraction speeds smaller than a critical value, $v_c = (B/\mu) (h_0^4/R^3\ell^3)$. 

The experimental data for the maximum displacement shown in Fig.~\ref{fig:fluide}(b) are also well described by the theory, even though it is slightly underestimated at low retraction speeds. For $\Sigma \gtrsim 1$, the dispersion of the data is lower, and the agreement with the theory is very good because, in this regime of high adhesion, the imposed displacement is geometrically limited by the ring's deformation, which, being inextensible, is close to its maximum value.

\textit{Conclusion---}In this work, we demonstrate the added value of the microstructure's flexibility. Our experiments show that the maximum adhesive force can be increased by up to a factor of 100 when a deformable ring is used instead of a rigid structure to contact a flat, rigid surface with a thin layer of viscous fluid. This enhancement of adhesion at low retraction speed $v$ is primarily due to the fact that the maximum adhesive force is proportional to $v$ in the rigid case, whereas it is proportional to the square-root of $v$ in the elastic case. The adhesion enhancement is thus particularly significant at sufficiently low retraction speeds. At large retraction speeds, the adhesion force is high and strongly deforms the elastic structure, whose local curvature in the wet region makes the maximum adhesive force smaller than in the rigid case. The nonlinear mechanical response of the Elastica structure thus provides a clear advantage for adhesion: their deformability can be used to enhance or impair the adhesion force compared to rigid systems. This highlights the potential of adjusting adhesion by tuning the elasticity of the structures, offering a powerful route for the design of advanced adhesive systems. This work therefore paves the way for further investigations, particularly into other Elastica geometries, to explore how structural shape influences adhesion and to elucidate the fundamental physical mechanisms governing this complex adhesive behavior.

\bibliographystyle{apsrev4-2}
\bibliography{bibli}

\clearpage

\section{Supplemental Material}
\setcounter{equation}{0}
\setcounter{figure}{0}
\setcounter{page}{1}
\renewcommand{\theequation}{S\arabic{equation}}
\renewcommand{\thefigure}{S\arabic{figure}}
\renewcommand{\thepage}{S\arabic{page}}

\subsection{Exact solution for the Elastica}

\begin{figure}[!b]
\centering
\includegraphics[width=0.8\columnwidth]{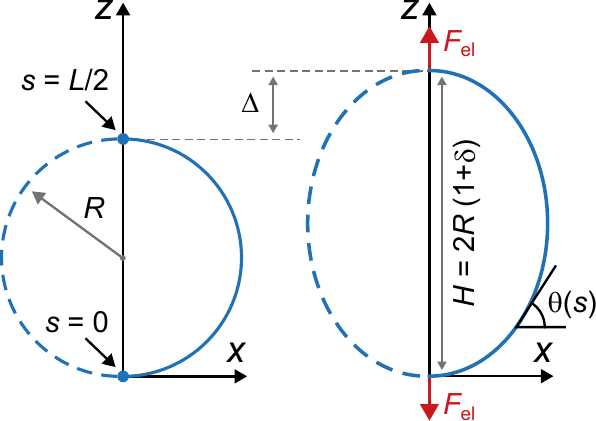}
\caption{Schematic of a circular elastic ring of length $L=2\pi R$ (left) deformed by a force $F_{\text{el}}$ (right). The applied force leads to a displacement $\Delta = 2R \delta$ of the point located at $s=L/2$.}
\label{fig-supp:schema}
\end{figure}

\textbf{Main equation.} The equation describing the elastic deformation of a circular ring of length $L=2\pi R$ subjected to a localized force $F_{\text{el}}$ acting along the vertical $z$-axis is given by
\begin{equation}
\label{sup-eq-elastica}
BW \frac{d^2 \theta(s)}{ds^2}+ \frac{F_{\text{el}}}{2} \cos \theta(s) =0,
\end{equation}
where $\theta(s)$ is the local angle between the tangent to the ring and the horizontal $x$-axis and $s$ the arclength varying between 0 and $L/2$, see Fig.~\ref{fig-supp:schema}. Due to symmetry, half of the system is considered, and consequently, half the applied force appears in Eq.~(\ref{sup-eq-elastica}). Therefore, in our case, the boundary conditions (BCs) are $\theta(0)=0$ and $\theta(L/2) = \pi$. Once $\theta(s)$ is known, the parametric equations that give the shape of the ring in Cartesian coordinates read as follows:
\begin{equation}
\label{sup-cartesian}
x(s) = \int_{0}^{s} \cos\theta(s') \, ds', \quad z(s) = \int_{0}^{s} \sin\theta(s')ds'. 
\end{equation}
Using the dimensionless variable $s = \bar{s}\, L/2$, so that $0 \le \bar{s} \le 1$, Eq.~(\ref{sup-eq-elastica}) becomes 
\begin{equation}
\label{sup-eq-elastica-adim}
\frac{d^2\theta}{d\bar{s}^2} + q^2 \cos\theta=0, \quad  q^2 = \frac{F_{\text{el}} L^2}{8BW} = \frac{\pi^3}{4} \bar{F}_{\text{el}},
\end{equation}
to be solved with the BCs $\theta(0)=0$ and $\theta(1)=\pi$. 

\textbf{Initial solution.} Initially, there is no applied force ($F_{\text{el}}=q=0$). The solution of Eq.~(\ref{sup-eq-elastica-adim}) satisfying the BCs is then given by
\begin{equation}
\label{sup-theta0}
\theta_0 = \pi \bar{s} = s/R, \quad q_0 = 0,
\end{equation}
which describes a circle with a radius $R$.
 
\textbf{Linear solution.} We expect the deformation to be small when the applied force is sufficiently small. The initial solution is therefore perturbed as follows
\begin{equation}
\label{sup-expansion-t-q}
\theta = \theta_0 + \epsilon^2 \theta_1, \quad q = q_0 + \epsilon q_1.
\end{equation}
Substituting these expansions into Eq.~(\ref{sup-eq-elastica-adim}), we obtain to the first order in $\epsilon$
\begin{equation}
\label{sup-theta1-eq}
\frac{d^2\theta_1}{d\bar{s}^2} + q_1^2 \cos\theta_0=0, \quad \theta_1(0)=\theta_1(1)=0.
\end{equation}
Using Eq.~(\ref{sup-theta0}), this last equation can be integrated twice, and by using the BCs, we obtain
\begin{equation}
\label{sup-theta1}
\theta_1(\bar{s}) =  \frac{q_1^2}{\pi^2}\left(\cos(\pi \bar{s}) + 2\bar{s}-1\right).
\end{equation}
The relationship between the applied force and the relative elongation $\delta= (H-2R)/2R$ is obtained using Eq.~(\ref{sup-cartesian}) to compute $H = z(L/2)$:
\begin{align}
\label{sup-H-temp}
H &= \frac{L}{2} \int_0^{1} \sin(\theta_0(\bar{s}) + \epsilon^2 \theta_1(\bar{s}))\, d\bar{s}, \\
  &= \frac{L}{2} \left[\int_0^{1} \sin \theta_0(\bar{s}) d\bar{s} + \epsilon^2 \int_0^{1}\theta_1(\bar{s}) \cos \theta_0(\bar{s}) d\bar{s} \right], \nonumber 
\end{align}
where we have used the expansion (\ref{sup-expansion-t-q}) for $\theta$. Using the expressions (\ref{sup-theta0}) and (\ref{sup-theta1}) for $\theta_0$ and $\theta_1$, the two remaining integrals can be computed explicitly to obtain
\begin{equation}
H = \frac{L}{2} \left[\frac{2}{\pi} + \frac{\pi^2-8}{2\pi^4}(\epsilon q_1)^2\right] = 2R + 2R \frac{\pi^2-8}{4\pi^3}q^2,
\end{equation}
where we have used $L=2\pi R$ and the expansion (\ref{sup-expansion-t-q}) for $q$. Consequently, we have
\begin{equation}
\delta= \frac{H-2R}{2R} = \frac{\pi^2-8}{4\pi^3}q^2 = \frac{\pi^2-8}{16}\bar{F}_{\text{el}}, 
\end{equation}
where we have used the relation (\ref{sup-eq-elastica-adim}) between $q$ and $\bar{F}_{\text{el}}$. We have thus obtained a linear relationship between the applied force and the relative elongation
\begin{equation}
    \bar{F}_{\text{el}} = \frac{16}{\pi^2-8} \, \delta \simeq 8.56\, \delta.
\end{equation}

\textbf{Nonlinear solution.} Eq.~(\ref{sup-eq-elastica-adim}) can be integrated once by multiplying it by the integrating factor $d\theta/d\bar{s}$:
\begin{equation}
\frac{1}{2}\left(\frac{d\theta}{d\bar{s}}\right)^2 + q^2 \sin \theta = q^2 C,
\end{equation}
where $C$ is an integration constant. This separable differential equation can be integrated side by side to obtain
\begin{equation}
F((\pi-2\theta)/4,-k^2)= -\frac{q}{k}(\bar{s}+\bar{s}_0),
\end{equation}
where we set $k^2 = 2/(C-1)$ and where $F(x,k^2) = \int (1-k^2 \sin^2x)^{-1/2} dx$ is the incomplete elliptic integral of the first kind~\footnote{F. W. J. Olver, D. W. Lozier, R. F. Boisvert, and C. W. Clark, eds., NIST Handbook of Mathematical Functions (Cambridge University Press, Cambridge, 2010). \label{ma-note}}. This relationship can be inverted to obtain an explicit expression for $\theta$:
\begin{equation}
\label{sup-theta-nonlinear-general}
2\theta(\bar{s}) = \pi + 4\, \text{am}\left(\frac{q}{k}(\bar{s}+\bar{s}_0),-k^2\right),
\end{equation}
where $\text{am}(x,k^2)$ is the Jacobi amplitude function and is the inverse of the incomplete elliptic integral of the first kind, i.e. $F(\text{am}(x,k^2),k^2) = x$ and $\text{am}(-x,k^2)=-\text{am}(x,k^2)$~\footref{ma-note}. The parameters $q$ and $\bar{s}_0$ are fixed by the boundary conditions $\theta(0)=0$ and $\theta(1)=\pi$. Using Eq.~(\ref{sup-theta-nonlinear-general}), we get
\begin{subequations}
\label{sup-BC-nonlinear}
\begin{align}
&\text{am}\left(\frac{q}{k}\bar{s}_0,-k^2\right) = -\frac{\pi}{4}, \\
& \text{am}\left(\frac{q}{k}(1+\bar{s}_0),-k^2\right) = \frac{\pi}{4}.
\end{align}
\end{subequations}
Since the Jacobi amplitude function is an odd function, we get $-\bar{s}_0 = 1+\bar{s}_0$, so that $\bar{s}_0 = -1/2$. The value of $q$ is then obtained by inverting one of the two relations (\ref{sup-BC-nonlinear}): 
\begin{equation}
\label{sup-q-F}
\frac{q}{2k} = F\left(\frac{\pi}{4},-k^2\right) \ \Rightarrow \ \bar{F}_{\text{el}}(k) = \frac{16 k^2}{\pi^3} F\left(\frac{\pi}{4},-k^2\right)^2,     
\end{equation}
where we used the relation (\ref{sup-eq-elastica-adim}) between $q$ and $\bar{F}_{\text{el}}$ and $F(x,k^2)$ is again the incomplete elliptic integral of the first kind. The final expression of $\theta$ reads as follows
\begin{equation}
\label{sup-theta-nonlinear}
2\theta(\bar{s},k) = \pi + 4\, \text{am}\left((2\bar{s}-1)F(\pi/4,-k^2),-k^2\right),
\end{equation}
where we have added the explicit dependence on $k$ which is still a free parameter at this stage. Eq.~(\ref{sup-q-F}) shows that $k$ fixes the force $\bar{F}_{\text{el}}$. We now show that $k$ also fixes the relative elongation, so that we obtain a parametric equation for the force-displacement curve where $k$ is the parameter. 

\begin{figure}[t]
\centering
\includegraphics[width=\columnwidth]{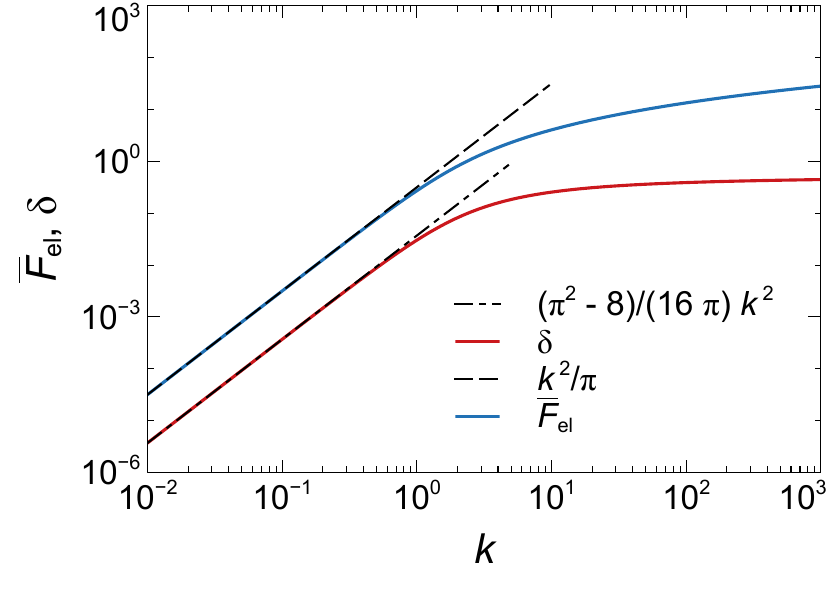}
\caption{Evolution of $\bar{F}_{\text{el}}$ and $\delta$, given respectively by Eqs.~(\ref{sup-q-F}) and (\ref{sup-delta-nonlinear}) as a function of $k$ together with their asymptotic expansions at low $k$.}
\label{fig-supp:graph-F-del-k}
\end{figure}

The relative elongation is again obtained from Eq.~(\ref{sup-cartesian}) to compute $H = z(L/2)$:
\begin{equation}
H(k) = \frac{L}{2} \int_0^1 \sin \theta(\bar{s},k) \, d\bar{s}.
\end{equation}
Since $\delta(k)=(H(k)-2R)/2R$ and $2R = L/\pi$, we get
\begin{equation}
\label{sup-delta-nonlinear}
\delta(k)= \frac{\pi}{2} \int_0^1 \sin \theta(\bar{s},k) \, d\bar{s} -1.
\end{equation}
In summary, by varying the value of $k$ from 0 to infinity, we obtain the parametric curve $(\delta(k),\bar{F}_{\text{el}}(k))$ where $\delta(k)$ is calculated from Eqs.~(\ref{sup-theta-nonlinear}) and (\ref{sup-delta-nonlinear}) and $\bar{F}_{\text{el}}(k)$ is calculated from Eq.~(\ref{sup-q-F}). This curve is shown in Fig.~\ref{fig:glue} of the main text. It features a linear growth of the force at small values of $\delta$ and a divergence at a finite value of $\delta$.

As shown in Fig.~\ref{fig-supp:graph-F-del-k}, both the force $\bar{F}_{\text{el}}$ and the relative elongation $\delta$ are increasing functions of $k$. When $k$ is small, both are proportional to $k^2$ so that $F$ varies linearly with $\delta$. For large values of $k$, the force continues to grow, while $\delta$ reaches its maximum value where $\theta = \pi/2$ for all $\bar{s}$, i.e. the ring is fully elongated. In this case, Eq.~(\ref{sup-delta-nonlinear}) gives $\delta_{\text{max}}=\pi/2 -1$. Since the force grows while $\delta$ remains fixed, the force diverges at $\delta = \delta_{\text{max}}$.

\begin{figure}[t]
\centering
\includegraphics[width=0.8\columnwidth]{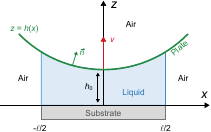}
\caption{Schematic of the system where a liquid of viscosity $\mu$ is placed between a rigid planar substrate and a rigid plate of arbitrary symmetric shape with $h(x)=h(-x)$.}
\label{fig-supp:schema-adhesion}
\end{figure}

\subsection{Viscous adhesion between a planar substrate and a rigid plate}

\textbf{General equations.} We consider a two-dimensional system where a thin layer of liquid with viscosity $\mu$ is placed between a rigid planar substrate and a rigid plate of arbitrary shape $z=h(x)$ symmetric with respect to $x=0$, which moves upward with a constant velocity $v$, see Fig.~\ref{fig-supp:schema-adhesion}. We assume that $h/\ell \ll 1$, where $\ell$ is the length of the wet region of the substrate, so that we can use the lubrication approximation. The Navier-Stokes equations in the plane ($x$,$z$) reduce to
\begin{equation}
    \label{lubri}
    \mu \frac{\partial^2 v_x}{\partial z^2}=\frac{\partial p}{\partial x}, \quad \frac{\partial p}{\partial z}=0, \quad \frac{\partial v_x}{\partial x}+\frac{\partial v_z}{\partial z}=0,
\end{equation}
where gravity has been neglected. These equations must be solved with the following BCs
\begin{subequations}
\begin{align}
    \label{BC-v-z=0}
    &v_x=0, \quad v_z=0 \quad &&\text{at}\quad z=0, \\
    \label{BC-v-z=h}
    &v_x=0,\quad v_z=v \quad &&\text{at}\quad z=h(x),\\
    \label{BC-p1}
    &\frac{\partial p}{\partial x}=0 \quad &&\text{at}\quad x=0, \\
    \label{BC-p2}
    &p=0 \quad &&\text{at}\quad x=\ell/2.
\end{align}
\end{subequations}
The BCs (\ref{BC-v-z=0}) and (\ref{BC-v-z=h}) are the non-slip BCs for the velocity field. The BCs~(\ref{BC-p1}) and (\ref{BC-p2}) take into account the symmetry of the system and that the liquid is in contact with air, whose pressure is set to 0. 

The second of Eqs.~(\ref{lubri}) shows that the pressure depends only on $x$, $p=p(x)$, so that the first of Eqs.~(\ref{lubri}) can be integrated twice. Using BCs (\ref{BC-v-z=0}) and (\ref{BC-v-z=h}) for $v_x$, we get
\begin{equation}
    \label{vx}
    v_x(x,z)=\frac{1}{2\mu}\frac{dp(x)}{dx}\,z\,[z-h(x)].
\end{equation}
The expression of $v_z$ is obtained by integrating the last of Eqs.~(\ref{lubri}) with the BC (\ref{BC-v-z=0}) for $v_z$
\begin{equation}
    \label{vz}
    v_z(x,z)=-\frac{z^3}{6\mu}\frac{d^2p(x)}{dx^2}+\frac{z^2}{4\mu}\frac{d}{dx}\left(h(x)\frac{dp(x)}{dx}\right).
\end{equation}
The last BC (\ref{BC-v-z=h}) for $v_z$ gives the following equation for the pressure
\begin{equation}
    12 \mu v = \frac{d}{dx}\left[h^3(x) \frac{dp(x)}{dx}\right].
\end{equation}
This equation can be integrated twice with the BCs~(\ref{BC-p1}) and (\ref{BC-p2}) to obtain
\begin{equation}
\label{p-eq}
  p(x) = 12 \mu v \int_{\ell/2}^x \frac{x'}{h^3(x')} dx'.
\end{equation}

Once the pressure is calculated, the velocities (\ref{vx}) and (\ref{vz}) are known and the force per unit width $W$ required to move the plate upward is obtained from the stress integrated along the surface of the plate 
\begin{subequations}
\begin{align}
    \label{FA-def}
    \frac{F_{\mu}}{W}&=\int_{-\ell/2}^{\ell/2} \tau_{nz}(x,z=h(x))\, \left[1+h'(x)^2\right]^{1/2}\, dx,\\
    \tau_{ij}&=-p\,\delta_{ij} + \mu\left(\frac{\partial v_i}{\partial x_j}+\frac{\partial v_j}{\partial x_i}\right),
\end{align}
\end{subequations}
where $h'(x) \equiv dh(x)/dx$, $\tau_{ij}$ is the stress tensor and $\vec{n} = \vec{\nabla}(z-h(x))/|\vec{\nabla}(z-h(x))|$ a unit vector normal to the plate whose explicit expression reads as
\begin{equation}
    \vec{n} = n_x \vec{e}_x + n_z \vec{e}_z = -\frac{h'(x)\, \vec{e}_x}{\sqrt{1+h'(x)^2}}  + \frac{\vec{e}_z}{\sqrt{1+h'(x)^2}}.
\end{equation}
The component of interest in the stress tensor is then given by
\begin{equation}
\label{taunz-general}
    \tau_{nz} = \vec{n} \cdot \overset{\leftrightarrow}{\tau} \cdot \vec{e}_z = n_x \tau_{xz} + n_z \tau_{zz},
\end{equation}
where
\begin{equation}
    \tau_{xz} = \mu\left(\frac{\partial v_x}{\partial z}+\frac{\partial v_z}{\partial x}\right), \quad \tau_{zz} = -p+2\mu \frac{\partial v_z}{\partial z}.
\end{equation}
The expression of $\tau_{nz}$ simplifies significantly in the lubrication approximation. Indeed, considering the following orders of magnitude $x\sim \ell$ and $z \sim h$, mass conservation (\ref{lubri}) yields $v_x \sim \ell v/h$. Taking into account $h/\ell \ll 1$, as already assumed above, we have
\begin{equation}
\label{orders-magnitude}
    n_x \tau_{xz} \sim \frac{\mu v}{h}, \quad n_z \tau_{zz} \sim -p + 2\frac{\mu v}{h}, \quad p\sim \frac{\mu v \ell^2}{h^3},
\end{equation}
where the estimation of $p$ comes from Eq.~(\ref{p-eq}). These relations (\ref{orders-magnitude}) show that $\tau_{nz} \simeq -p$ provided $(h/\ell)^2 \ll 1$. Consequently, at the same order in the small parameter $h/\ell$, the force required to lift the plate is given by
\begin{equation}
    \label{FA-eq-temp}
    \frac{F_{\mu}}{W}=-2\int_{0}^{\ell/2} p(x)\, dx,
\end{equation}
where the fact that $p$ is an even function has been taken into account. Substituting the expression (\ref{p-eq}) of $p$ in Eq.~(\ref{FA-eq-temp}) and integrating once by parts, we finally obtain
\begin{equation}
    \label{FA-eq}
    \frac{F_{\mu}}{W}=24 \mu v \int_{0}^{\ell/2}\frac{x^2}{h^3(x)}\, dx.
\end{equation}
Therefore, once the shape of the plate is given, the force $F_{\mu}$ can be computed.

\textbf{Rigid planar plate.} When the moving plate is planar, we have $h(x)=h_0$ and Eq.~(\ref{FA-eq}) leads to
\begin{equation}
    F_{\mu}^{\text{flat}} = \frac{\mu v W \ell^3}{h_0^3}.
\end{equation}
Since both the wet region of the substrate of length $\ell$ and the height $h_0$ vary during the retraction, it is useful to eliminate $\ell$ in favor of the volume $\Omega = W \ell h_0$ of fluid, which remains constant during this process. We thus finally obtain
\begin{equation}
\label{FA-eq-planar}
    F_{\mu}^{\text{flat}} = \frac{\mu v\, \Omega^3}{W^2 h_0^6}.
\end{equation}

\begin{figure}[t]
\centering
\includegraphics[width=\columnwidth]{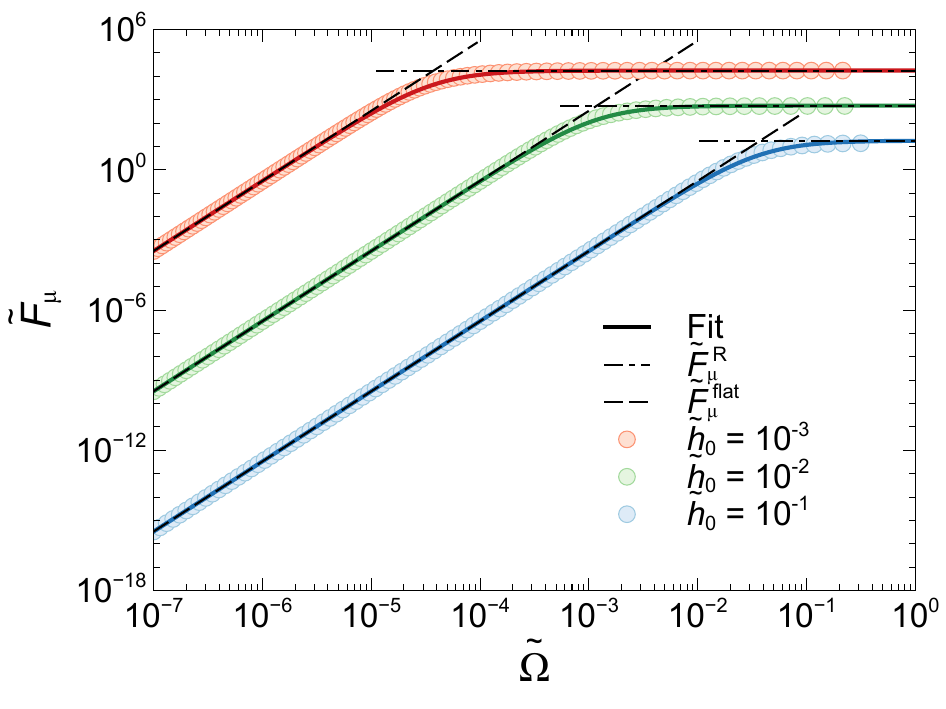}
\caption{Evolution of $\tilde{F}_{\mu}$ as a function of $\tilde{\Omega}$, both given by Eq.~(\ref{Fmu-Om-adim}), for three values of $\tilde{h}_0$. The symbols are obtained from numerical computations. The asymptotics expressions (\ref{Fmu-asymp}) are also shown together with the fit (\ref{Fmu-Fit}).}
\label{fig-supp:adhesion-force}
\end{figure}

\textbf{Rigid plate with constant curvature.} When the moving plate is a circular arc with a radius $R$, its equation is given by
\begin{equation}
    h(x)=R+ h_0-\sqrt{R^2-x^2}.
\end{equation}
The adhesion force (\ref{FA-eq}) and the volume of liquid are then given by
\begin{align}
    F_{\mu} &= 24 \mu v W \int_{0}^{\ell/2}\frac{x^2}{h^3(x)}\, dx, \\
    \Omega &=2 W \int_0^{\ell/2} h(x)\, dx. 
\end{align}
Using the change of variables 
\begin{subequations}
\begin{align}
    x &= R\, \tilde{x}, \quad h = R\, \tilde{h}, \quad h_0 = R\, \tilde{h}_0, \quad \ell =2R\, \tilde{\ell}, \\
    F_{\mu} &= 24 \mu v W\, \tilde{F}_{\mu}, \quad \Omega = 2 WR^2\, \tilde{\Omega},
\end{align}
\end{subequations}
we have
\begin{subequations}
\begin{align}
\label{Fmu-Om-adim}
    \tilde{F}_{\mu} &= \int_{0}^{\tilde{\ell}}\frac{\tilde{x}^2}{\tilde{h}^3(\tilde{x})}\, d\tilde{x}, \quad \tilde{\Omega} = \int_0^{\tilde{\ell}} \tilde{h}(\tilde{x})\, d\tilde{x},\\
    \label{hbar}
    \tilde{h}(\tilde{x}) &= 1+ \tilde{h}_0-\sqrt{1-\tilde{x}^2}.
\end{align}
\end{subequations}
The dimensionless force, $\tilde{F}_{\mu}$, and volume, $\tilde{\Omega}$, are both functions of $\tilde{h}_0$ and $\tilde{\ell}$. Equations (\ref{Fmu-Om-adim}) are thus the parametric equations for $\tilde{F}_{\mu}$ as a function of $\tilde{\Omega}$ with $\tilde{\ell}$ as a parameter. Their expressions can be computed analytically. However, they are cumbersome and will not be written here. The resulting curve obtained numerically still depends on $\tilde{h}_0$ as shown in Fig.~\ref{fig-supp:adhesion-force}. This figure shows that the adhesion force can be described by the following asymptotic expressions when $\tilde{\Omega}$ is small or sufficiently large: 
\begin{subequations}
\label{Fmu-asymp}
\begin{align}
\label{Fmu-asymp1}
    \tilde{F}_{\mu} &= \tilde{F}_{\mu}^{\text{flat}} = \frac{\tilde{\Omega}^3}{3 \tilde{h}_0^6}  &\text{when}&  &\tilde{\Omega} &\ll \tilde{h}_0^{3/2}, \\
\label{Fmu-asymp2}
    \tilde{F}_{\mu} &= \tilde{F}_{\mu}^{\text{R}} = \frac{\pi}{4\sqrt{2} \tilde{h}_0^{3/2}}  &\text{when}&  &\tilde{\Omega} &\gg \tilde{h}_0^{3/2}.
\end{align}
\end{subequations}
The expression (\ref{Fmu-asymp1}) comes from the fact that this regime corresponds to large $R$ where $\tilde{\ell} \ll 1$ and $\tilde{x} \ll 1$ so that $\tilde{h}(\tilde{x}) \simeq \tilde{h}_0$ in Eq.~(\ref{hbar}). The expression (\ref{Fmu-asymp2}) is obtained by considering $\tilde{\ell} =1$ and the limit $\tilde{h}_0 \ll 1$ in the expression of $\tilde{F}_{\mu}$.

Finally, the expression for the adhesion force is well approximated by the following fit
\begin{equation}
    \label{Fmu-Fit}
    \tilde{F}_{\mu} \approx \tilde{F}_{\mu}^{\text{flat}}\left[1+ \left(\tilde{F}_{\mu}^{\text{flat}}/\tilde{F}_{\mu}^{\text{R}}\right)^{1/2}\right]^{-2}.
\end{equation}
The dimensional version of this relation is simply obtained by removing the tildes, using the expression (\ref{FA-eq-planar}) of $F_{\mu}^{\text{flat}}$ and $F_{\mu}^{\text{R}}=6\pi/(\sqrt 2) \mu v W (R/h_0)^{3/2}$.

\begin{figure}[t!]
\centering
\includegraphics[width=\columnwidth]{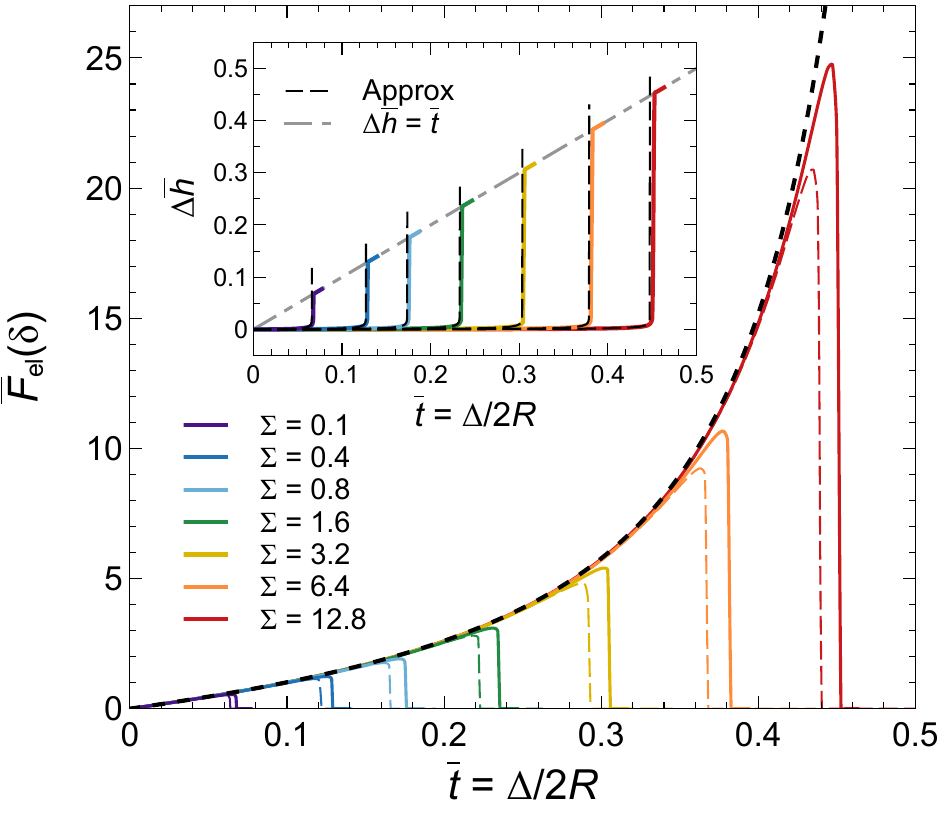}
\caption{Evolution of the rescaled elastic force $\bar{F}_{\text{el}}$ as a function of $\bar{t}=\Delta/2R$ obtained by solving numerically Eq.~(\ref{equ:nonlinear_ODE-sup}) with $\bar{h}_{0}=0.01$, and several values of $\Sigma$. The solid curves correspond to $\Lambda =0$ whereas the dashed ones correspond to $\Lambda =0.1$. The dotted black curve represents the elastic force for glued rings, see Eq.~(\ref{eq:fin}) of the main text. The force reaches a maximum value $\bar{F}_{\text{max}}$, which increases with $\Sigma$, before to vanish when detachment occurs. Inset: Evolution of the rescaled increase of fluid thickness, $\Delta \bar{h} = (h-h_0)/2R$, as a function of the rescaled imposed displacement $\Delta/2R = \bar{t}$. The dashed curves show the approximate solution (\ref{approx-solution-h-sup}). The imposed displacement at which detachment occurs, $\Delta_{\text{max}}/2R$, increases with the rescaled retraction speed $\Sigma$. The dashed-dotted curve shows the evolution of $\Delta \bar{h}$ in the absence of adhesion, i.e. $h(t)=h_0 + vt$.
}
\label{fig:model}
\end{figure}

\subsection{Coupling fluid-structure}

To model the elasto-viscous adhesion, the viscous force (\ref{Fmu-Fit}) should be combined with the mechanical response of the flexible ring given by Eq.~(\ref{eq:fin}) of the main text. As shown in the main text, the dynamics is governed by $F_{\mu}(h(t),R) = F_{\text{el}}(\delta(t))$ together with $2R\delta(t) = h_0+vt - h(t)$. We thus obtain
\begin{subequations}
\label{equ:nonlinear_ODE-sup}
\begin{align}
\label{equ:nonlinear_ODE-sup-1}
&\frac{\Sigma}{\bar h^6 \left[1 + \Lambda \,\bar{h}^{-9/4}\right]^2}
\frac{d\bar{h}}{d\bar{t}}
=\bar{F}_{\text{el}}\left[\bar{h}_{0}(1-\bar{h}) + \bar{t}\right], \\
\label{equ:nonlinear_ODE-sup-2}
&\Sigma = \frac{\mu v R \Omega^3}{\pi BW^3 h_0^5}, \quad \Lambda = \frac{2^{1/4}}{\sqrt{6\pi}}\, \left[\frac{\Omega}{W h_0^{3/2} R^{1/2}}\right]^{3/2},
\end{align}
\end{subequations}
where $\bar h=h/h_0$, $\bar h_0=h_0/2R$, $\bar t= vt/2R$ and both forces $F_{\mu}$ and $F_{\text{el}}$ have been rescaled as in the main text, i.e. $\bar{F}_{i}=2 F_{i} R^2/(\pi B W)$ with $i=\mu$ or el. Figure~\ref{fig:model} shows that the adhesion force follows the elastic response of the glued rings up to the sudden drop of force during the very short interval of time in which detachment occurs. It also shows that the fluid thickness remains almost constant before suddenly increasing at $\bar{t}_{\text{max}}=\Delta_{\text{max}}/2R$. An asymptotic solution can thus be obtained by setting $\bar h=1$ in the argument of $\bar{F}_{\text{el}}$ and considering the limit $\Lambda \to 0$. Equation~(\ref{equ:nonlinear_ODE-sup-1}) then reduces to
\begin{equation}
    \frac{\Sigma}{\bar h^6} \frac{d\bar{h}}{d\bar{t}} =\bar{F}_{\text{el}}(\bar{t}),
\end{equation}
which can be integrated with the initial condition $\bar{h}(0)=1$ to obtain
\begin{equation}
\label{approx-solution-h-sup}
    \bar{h}(\bar{t}) = {\Sigma}^{1/5} \left[\Sigma - 5\, G_{\text{el}}(\bar{t}) \right]^{-1/5},
\end{equation}
where 
\begin{align}
   G_{\text{el}}(\bar{t}) &= \int_0^{\bar{t}} \bar{F}_{\text{el}}(t') dt', \nonumber \\
   &=  \frac{1}{2} \bar{t}^{\, 2} \left[\alpha+\frac{8 \beta}{(\pi -2) (\pi -2 (\bar{t}+1))}\right],
\end{align}
with $\alpha = 3.86$ and $\beta = 0.765$ and where we have used 
\begin{equation}
    \bar{F}_{\text{el}}(x)  \simeq \alpha\, x + \frac{2\beta (\pi-2)}{[\pi-2(1+x)]^2} - \frac{2\beta}{\pi-2},
\label{eq:fin-sup}
\end{equation}
see Eq.~(\ref{eq:fin}) of the main text. As shown in the inset of Fig.~\ref{fig:model}, the approximate solution (\ref{approx-solution-h-sup}) describes well the numerical solutions of Eq.~(\ref{equ:nonlinear_ODE-sup}).

The approximate solution (\ref{approx-solution-h-sup}) implies that the detachment occurs when
\begin{equation}
\label{detachment-cond}
    G_{\text{el}}(\bar{t}_{\text{max}}) = \Sigma/5.
\end{equation}
The function $G_{\text{el}}$ has the following asymptotic behaviors:
\begin{subequations}
\begin{align}
\label{Gel-small}
    G_{\text{el}}(x) &= 4.28\, x^2, \qquad x \ll 1 \\
\label{Gel-large}
    G_{\text{el}}(x) &= 0.437\, \left(\frac{\pi}{2}-1 - x \right)^{-1}, \qquad x \simeq \frac{\pi}{2}-1.
\end{align}
\end{subequations}
When $\Sigma \ll 1$, Eq.~(\ref{detachment-cond}) imposes that $G_{\text{el}}$ must be small, which occurs when $\bar{t}_{\text{max}} \ll 1$. Therefore, solving Eq.~(\ref{detachment-cond}) using the development (\ref{Gel-small}) for $G_{\text{el}}$ leads to
\begin{equation}
    \bar{t}_{\text{max}} = \frac{\Delta_{\text{max}}}{2R} = 0.216\, \Sigma^{1/2}. 
\end{equation}
The maximum adhesion force is expressed as:
\begin{equation}
\label{Fmax-small-sig-supp}
    \bar{F}_{\text{max}}= \bar{F}_{\text{el}}(\bar{t}_{\text{max}}) = 1.85\, \Sigma^{1/2}, \quad \Sigma \ll 1,
\end{equation}
where $\bar{F}_{\text{el}}$ has been expanded near $\Sigma=0$. When $\Sigma \gg 1$, Eq.~(\ref{detachment-cond}) imposes that $G_{\text{el}}$ must be large, which occurs when $\bar{t}_{\text{max}} \simeq \pi/2-1$. Therefore, solving Eq.~(\ref{detachment-cond}) using the development (\ref{Gel-large}) for $G_{\text{el}}$ leads to
\begin{equation}
    \bar{t}_{\text{max}} = \frac{\pi}{2}-1-\frac{2.18}{\Sigma}. 
\end{equation}
The maximum adhesion force is expressed as:
\begin{equation}
    \bar{F}_{\text{max}}= \bar{F}_{\text{el}}(\bar{t}_{\text{max}}) = 0.092\, \Sigma^{2}, \quad \Sigma \gg 1,
\end{equation}
where $\bar{F}_{\text{el}}$ has been expanded for $\Sigma \gg 1$.

\subsection{Adhesion enhancement}

\begin{figure}[t!]
\centering
\includegraphics[width=\columnwidth]{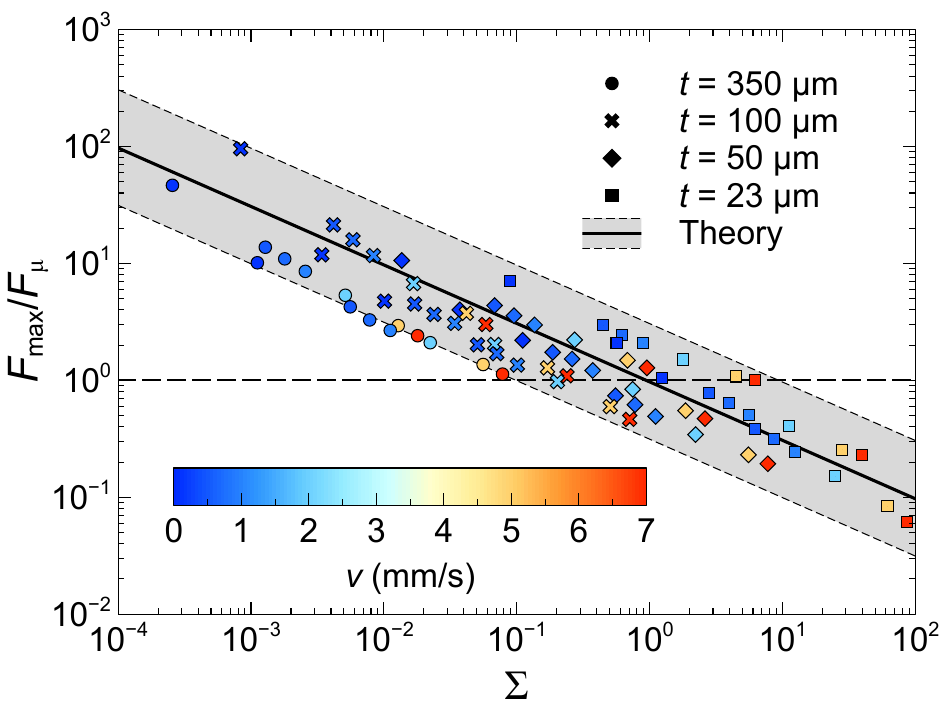}
\caption{Evolution of the ratio $F_{\text{max}}/F_{\mu}$ as a function of the rescaled retraction speed $\Sigma$. The shaded area corresponds to the region spanned by the theoretical expression (\ref{Fmax-over-Fmu}) when $\bar{h}_{0}$ and $\Lambda$ vary withing their experimental range, i.e. $0.010 < \bar{h}_{0} < 0.039$ and $2.3 < \Lambda < 4.2$.
}
\label{fig:enhancement}
\end{figure}

To assess the impact of the deformability of the rings on the adhesion strength, we divide the maximum adhesion force measured experimentally for flexible rings, $F_{\text{max}}$, by the maximum adhesive force that applies to rigid rings, $F_{\mu}$ given by Eq.~(\ref{Fmu-Fit}). Figure~\ref{fig:enhancement} shows that the adhesion strength is enhanced ($F_{\text{max}}/F_{\mu}\gtrsim 1$) when the rescaled retraction speed is roughly smaller than 1 ($\Sigma\lesssim 1$). Using the expression (\ref{Fmax-small-sig-supp}) of $F_{\text{max}}$ and the expression (\ref{Fmu-Fit}) $F_{\mu}$, we get
\begin{equation}
\label{Fmax-over-Fmu}
\frac{F_{\text{max}}}{F_{\mu}} = 2.9\, \bar{h}_{0} (1+\Lambda)^2\, \Sigma^{-1/2}.
\end{equation}
This expression agrees well with the data, as shown in Fig.~\ref{fig:enhancement} where the shaded area corresponds to the region spanned by Eq.~(\ref{Fmax-over-Fmu}) when $\bar{h}_{0}$ and $\Lambda$ vary within their experimental range.

\end{document}